\documentstyle[12pt]{report}

\newcommand{\fs}{\footnotesize}
\newcommand{\bsp}{\begin{sloppypar}}
\newcommand{\esp}{\end{sloppypar}}
\newcommand{\ds}{\displaystyle}
\newcommand{\la}[1]{\label{#1}}
\newcommand{\re}[1]{\ (\ref{#1})}
\newcommand{\nn}{\nonumber}
\newcommand{\ed}{\end{document}}
\newcommand{\be}{\begin{equation}}
\newcommand{\ee}{\end{equation}}
\newcommand{\ba}{\begin{eqnarray}}
\newcommand{\ea}{\end{eqnarray}}
\newcommand{\bb}{\begin{thebibliography}}
\newcommand{\eb}{\end{thebibliography}}
\newcommand{\bi}[1]{\bibitem{#1}}
\newcommand{\ct}[1]{\cite{#1}}

\newcommand{\ra}{\rightarrow}

\newcommand{\ap}{approximation}

\newcommand{\bm}{bag model}

\newcommand{\cd}{condensate}
\newcommand{\cf}{confinement}
\newcommand{\cnt}{constant}
\newcommand{\cef}{coefficient}

\newcommand{\con}{contribution}

\newcommand{\dc}{distance}

\newcommand{\ef}{effect}

\newcommand{\ex}{experiment}

\newcommand{\fw}{in the framework of}
\newcommand{\fq}{frequency}
\newcommand{\fu}{function}

\newcommand{\hn}{hadron}

\newcommand{\s}{instanton}
\newcommand{\sld}{instanton liquid}
\newcommand{\ia}{interaction}
\newcommand{\il}{integral}

\newcommand{\La}{Lagrangian}

\newcommand{\np}{nonperturbative}

\newcommand{\op}{operator}

\newcommand{\pr}{parameter}

\newcommand{\ptb}{perturbative}

\newcommand{\ph}{phenomenolog}

\newcommand{\pto}{proportional to}

\newcommand{\ps}{pseudoscalar}
\newcommand{\va}{the {\bf QCD} vacuum}

\newcommand{\qn}{quantum number}

\newcommand{\q}{quark}
\newcommand{\qm}{quark model}

\newcommand{\spl}{splitting}

\newcommand{\si}{singlet}

\newcommand{\str}{structure}

\newcommand{\ti}{there is}
\newcommand{\tsi}{takes into account}

\newcommand{\ten}{taken into account}

\newcommand{\vm}{vacuum}
\newcommand{\vq}{valence quark}

\newcommand{\w}{where}
\newcommand{\zfm}{zero fermionic mode}
\newcommand{\zm}{zero  mode}
\begin{document}
{\hfill     preprint JINR E2-93-117}
{\hfill     Z. Phys. C to be published}
\begin{center}
{\bf FOUR-QUARK STATES AND NUCLEON-ANTINUCLEON ANNIHILATION
WITHIN THE QUARK MODEL WITH INSTANTON INDUCED INTERACTION}\\[5mm]
A. E. DOROKHOV
\footnote{\large {
E-mail:dorokhov@thsun1.jinr.dubna.su}}\\
{\small \it Joint Institute for Nuclear Research,} \\
{\small \it Bogoliubov Theoretical Laboratory,} \\
{\small \it 141 980, Dubna, Moscow Region, Russia}\\[3mm]
N. I. KOCHELEV\\
{\small \it Joint Institute for Nuclear Research,} \\
{\small \it 141 980, Dubna, Moscow Region, Russia}\\[3mm]
  Yu. A. ZUBOV\\
{\small \it High Energy Physics Institute,Kazakh Academy of Sciences  } \\
{\small \it SU-480082 Alma~-~Ata, Republic of Kazakhstan}\\[3mm]

{ Abstract} \\
\end{center}
The spectrum of $q^2\bar q^2, J^p=0^+,2^+$ mesons is discussed.
We have shown that due to \s\ - induced forces the physical states are strong
mixtures of the $SU_f(3)$
group basis states. The cross-sections for annihilation of
the $N\bar N$ system
into $(q^2\bar q^2)(q\bar q)$ mesons are obtained. The $a_0(980)$ meson is
considered as $q^2\bar q^2$ meson consisting of $9_f$ and $36_f$ plets.
The branchings are also predicted for the cross-sections for production of
the $a_0(980)$ and tensor $q^2\bar q^2$ mesons in $N\bar N$ annihilation.
\vspace{1.0cm}

\begin{sloppypar}

{\bf INTRODUCTION} \\

\ \ \ \  Four-quark mesons were first considered in papers
\ct{Jaf}, \ct{Wong} in the MIT version of the bag model. From this
consideration it followed in particular that the known $a_0(980),\  f_0(975)$
mesons may be assigned to the lightest $q^2\bar q^2$ states.
Now there exist some additional experimental indications of multiquark meson
states.
Let us mention the most clear of them. In the experiments on
vector meson VV' production in $\gamma \gamma $ scattering there are
observed resonance - like signals which are probably of the four-quark
nature \ct{Exp}. At the Serpukhov facility, in \hn -\hn\ scattering
the extraordinary  C-resonance was discovered in the $\phi \pi $
system with  mass
$\sim $1,5 GeV and quantum numbers (I=1, $J^{pc}=1^{--}$) \ct{Bit}.
Later on, the GAMS group reported on  the resonance structure in the
$\eta \pi $ channel \ct{Ald} with exotic
quantum numbers (I=1, $J^{pc}=1^{-+}$) which is probably a G-partner of
the C-resonance.

Rich material for the $q^2\bar q^2$ meson spectroscopy can be extracted
from $N\bar N \ra (q^2 \bar q^2) (q\bar q) \ra mesons$
reactions \ct{NN4q}. In paper \ct{Br}
the resonance observed in $\bar pn \ra \pi ^-X_0(1480) \ra 3\pi ^- 2\pi ^+$
reaction is analyzed.
The decay of  $X_0(1480)$ goes mainly through the $\rho ^0 \rho ^0$ system.
This resonance possibly also contributes to $\gamma \gamma \ra \rho ^0 \rho
^0$ reaction and was observed by the TACCO,  CELLO and JADE \ct{Exp} groups.
Recently the ASTERIX collaboration has presented the data on the annihilation
of $p \bar p$ into the $\pi ^+ \pi ^- \pi ^0$ system \ct{May}  where
the resonance in the $\pi ^+ \pi ^- $ mass spectrum  with a mass $\sim $
1565 MeV
and a width  170 MeV is observed. The Crystal Barrel collaboration in LEAR
has observed a tensor meson with a width $\sim 120 MeV$ in the
$\bar pp \ra \pi ^0X_2(1515) \ra \pi ^0 \pi ^0 \pi ^0$ reaction \ct{Ak}.
In many papers  the E(1420) meson  with a width $\sim
$ 60 MeV  observed for example in $\bar pp \ra (K^{\pm }K_s \pi
^{\mp})\pi ^+ \pi ^-$ is widely discussed
\ct{Duch}, \ct{Bai}. The decay of the E-meson goes
mainly through $a_0(980) \pi\ (a_0(980) \ra \bar KK)$. So, if $a_0(980)$ is
really a four-quark state, then it is natural to suppose that
the E-meson is also a four-quark state. The global discussion on all four -
\q\ system has been considered in \ct{KK4q}.

All these experimental data need a theoretical analysis to answer the
question: are these really the states that may be described as $q^2\bar q^2$
states? It is necessary to calculate the
$q^2\bar q^2$ meson mass spectrum and their wave functions
in the quark model where the mixing of hadron states is taken into account
and investigate their decay properties.

This paper is concerned mainly with phenomenology: the spectrum of
$q^2\bar q^2$ mesons (based on \ct{DKZ4Q}), their couplings and the
possibility that some $N\bar N$ channels are related with these exotic
mesons.
In the first section of this paper we review the quark model with
the interaction induced by the vacuum of Quantum Chromodynamics (QCD).
In the second section, in the framework of this model we present the mass
spectrum of $q^2\bar q^2$ mesons, their couplings and the basis of
physical states.

Sections 3, 4 and 5 contain the application of the results obtained.
Our primary predictions concern signals indicating the
possible four - quark nature of scalar (in particular, the $a_0(980)-$ meson),
vector and tensor mesons from their production in $N\bar N$ reaction.
This analysis is quite general. It is based only on the group - theoretical
structure of this reaction and does not use a concrete dynamical model.
As phenomenological parameters we use the mass of four - quark states
predicted in our model and strong decay constants fitted from \ex .
We expect the most bright manifestation of the four - \q\ states in the
reaction $N\bar N\to(4q)$ {\it tensor meson} $+\ (q\bar q)$
{\it vector meson} discussed
in Sect. 5. Specifically,  in the reaction $n\bar p\to\rho^-\rho^-\rho^+$,
we predict the appearance of the exotic double charged $E^{--}_{\pi \pi} $
tensor
meson in the invariant mass spectrum of the $\rho^-\rho^-$ system \re{E--}.
This prediction is made for the first time. Another
thing is that if the hypothesis of four - \q\ origin of the $a_0(980)$
is right, then from the observation of the $a_0(980)$ in
$p\bar p\to a_0(980)\pi_0$ (Crystal Barel, CELLO) \ct{AR,CB} and the
estimation \re{y24} given below it follows that one should expect the intense
yield of the exotic $E^{--}$ in the reaction $n\bar p\to E^{--}\rho^{++}$.

\chapter{Quark model with QCD vacuum - induced interaction}
    In \ct{DK} the quark bag model has been formulated with emphasis on
the quark-QCD vacuum field interaction.
A crucial fact allowing the construction of a realistic model of hadrons is
that \ti\
the physical medium surrounding a bag populated by collective
intensive vacuum fluctuations. The interaction of fields inside the bag with
external \vm\ fields completely changes the \str\ of the standard bag model.
Within this approach the bag surface is determined self-consistently by
minimizing the total energy of the \hn\ as a system of Dirac \q s in
an external vacuum field.

Let us consider the hierarchy of fields in the QCD \vm\ and the
bag-vacuum system. As it is known, \va\ has a complicated \str .
Conventionally the \np\ fields can be divided in two parts: a short-wave
component
that provides the helicity sensitive \ia\ of \q s at small \dc s and a
long-wave
component
that gives the \cf . In the framework of the \sld\ model\ct{Shil,DPil}
the first part is connected with a fine - granular \vm\ \str\ where the
one-\s\ \ia\
with \ef ive size $\rho_c\sim1.5\div2\ GeV^{-1}$ dominates. The second
component is connected with the long-wave collective excitations of the
\sld\ with wave length $\lambda\approx R_{hadron}\approx R_{conf}$, \w\
$R\approx3\rho_c$ is an average \dc\ between \s s  and
$R_{conf}\approx5\div6\ GeV^{-1}$ is a \cf\ radius.

The main assumption of our model \ct{DK}  is that \va\ is almost
not destroyed by color fields inside the \hn\ and the \ia\ of \q s and
gluons localized in the bag with \vm\ fields defines the \hn\ \str .
We shall regard the bag and fields localized in it as immersed in
the physical \s\ \vm , and we shall suggest that \q s give rise to
(practically) no disturbtion of
the local properties of this medium. This assumption is analogous to the
QCD sum rule (QCD SR) one. In the latter case as a probe, i.e. a nonlocal
object selecting the lowest \hn\ states, the correlator of \hn\ currents
is used like the bag in our \qm . Here, one also suggests that
local sources do not perturb the properties of physical \vm , the sizes
of \q\ and gluon condensates.

There are three different length scales of fluctuations in the bag-\vm\
system:  small-size \s\ fluctuations with characteristic frequency
$\epsilon_{\ds i}\sim1/\rho_c$,
fluctuations of fields localized inside the bag with \fq\
$\epsilon_q\sim \omega_{\ds q}/R_{bag}$,
and long-wave \vm\ fluctuations with $\epsilon_{vac}\sim1/R_{conf}.$
For low-lying excitations of \q s in the bag,
the factorization of small, intermediate and large \dc s occurs:
$1/\rho_c>>\omega_{\ds q}/R_{bag}>>1/R_{conf}$,
and the use of the \ef ive \La\ technique is correct\ct{DKZPi}. For the scale
$r\le\ \rho_c$ the main \ef\ is the \ia\ related with tunnelling
due to the \s ,
and at the scale $r\sim R_{bag}$ it is the confinement of \q s in the bag.
With this hierarchy of
\ia s we can regard that the \ia\ of \q s in the \hn\ mainly develops
on one \s\ on the background of external \vm\ medium.

Long-wave \vm\ fields $Q(x),{\cal A}^a_\mu(x)$ satisfy the classical
Yang-Mills equations (acting on the state of physical \vm
$\mid \underline 0>$)
\ba
&&(i\hat \nabla-m_i)Q^i(x)\mid\underline{0}>=0, \\ \relax
&&\nabla^{\mu}_{ab}G^b_{\mu\nu}(x)\mid\underline{0}>=g
\bar Q^i(x){\lambda^a\over2}\gamma_\nu Q^i(x)\mid\underline{0}>, \nn\\
&&G^{a}_{\mu\nu}=\partial_\mu {\cal A}^a_\nu-\partial_\nu {\cal A}^a_\mu+
gf^{abc}{\cal A}^b_\mu {\cal A}^c_\nu .    \nn
\la{17.3"}\ea
Their solutions are parametrized by \si\ renormalization-invariant averages
\ct{SVZ}:
\ba
&&<\underline{0}\mid{\alpha_{\ds s}\over2\pi}:G^a_{\mu\nu}(0)G^{a\mu\nu}(0):
\mid\underline{0}>
\approx0.012\ GeV^4,\\
&&<\underline{0}\mid\alpha_{\ds s}^{4/9}:\bar Q(0)Q(0):\mid\underline{0}>
\approx-(250\ MeV)^3,\ ...,\nn
\la{13.1}\ea
where the sizes of \cd s are determined \ph ically within the
current algebra and QCD SR. Dots stands for the \op s of higher dimensions
and the normal ordering of operators with respect to \ptb\ \vm\
state $\mid0>$ is implied.

	To describe the \ia\ of \vq s with long-wave \vm\ fields, we make
the substitution
\be
q(x)\Theta_V(x)\rightarrow q(x)\Theta_V(x)+Q(x),\ \ \ \ \
A^a_\mu(x)\Theta_V(x)\rightarrow A^a_\mu(x)\Theta_V(x)+{\cal  A}^a_\mu(x)
\la{VT}\ee
in the \bm\ \La\
\be
{\cal L}^{QCD}\Theta_V(x)\rightarrow{\cal L}^{QCD}\Theta_V(x)+
\Delta{\cal L}^{vac},
\la{LB}\ee
where
\ba
\Delta{\cal L}^{vac}=&&[\bar q(x)\Theta_V(x)]({i\over2}
\stackrel{\leftrightarrow}{\hat\partial})Q(x)+
\bar
Q(x)({i\over2}\stackrel{\leftrightarrow}{\hat\partial})[q(x)\Theta_V(x)]+\nn\\
&&+g\bar q(x)\gamma^\mu{\lambda^a\over2} q(x){\cal A}^a_\mu(x)\Theta_V(x),
\la{14.15'}\ea
$Q$ and $\bar Q$ are anticommuting \vm\ \q\ fields,
${\cal A}^a_\mu(x)$ is an external gauge field. Localized field components
$q(x)$ and $A(x)$ are approximated by the solutions of the \bm\ equations
in the spherical cavity approximation \ct{PNB,MIT1}.

The \ia\ with the external long-wave \vm\ field \re{14.15'} results in an
additional energy increasing with bag size\ct{DK}.
As a consequence, there occurs a situation when a further growth becomes
impossible
(that is, large fluctuations of the bag size are strongly suppressed). Thus,
within our model the bag stability
is due to the \ia\ of bag fields with \vm\ fields.

The \ia\ of \q s with the short-wave component of \vm\ fields, small-size
\s s, is approximated by the \ef ive 't Hooft \La\ \ct{tH,SVZi}
that in the \sld\ model is \ct{Shil,NK}:
\be
\Delta{\cal L}^{inst}= \sum_{i>j}^{i=u,d,s}n_c(-k')^2\{\bar q_{iR}q_{iL}\bar
q_{jR}q_{jL}
[1+{3\over32}\lambda^{a}_{\ds i}\cdot\lambda^{a}_{j}
(1-{3\over4}\sigma_{\mu\nu}^{\ds i}\cdot\sigma_{\mu\nu}^{j})]+
(R\leftrightarrow L)\}
\la{12.7}\ee
where the \cnt\
\[
k'=\frac{4\pi\rho_C^3}{3}\frac{\pi}{(m_*\rho_C)}
\]
characterizes the \ia\ strength of a \q\  with an \s\ and is \pto\ the \s\
volume,
$n_c$ is the \s\ density in the \vm\ related to the \vm\ energy density by
$B_{QCD}\approx 2n_c$,
$ \rho_c$ is an \ef ive size of the \s\ in \va ,
$q_{R,L}=(1\pm\gamma_{\ds5})q/2$,
$m^*=-{\ds2\over\ds3}\pi^2\rho^2_c<~\underline{0}\mid
\bar QQ\mid~\underline{0}>$ is the \ef ive mass
of the \q\ with zero current mass in the physical \vm\ \ct{SVZi} ,
$<\underline{0}\mid\bar Q_{\ds i}Q_{\ds i}\mid\underline{0}>$ is a \q\ \cd .
An \ef ive mass \tsi\ long-range field correlations in the \s\ \vm .
The term $(R\leftrightarrow L)$ in \re{12.7} corresponds to the \ia\ through
an anti-\s . In addition, averaging over
\s\ space positions and orientations in color space is implicitly supposed.
The averaging over
collective variables provides the translation and gauge invariance of \s\ \ia .
The \La\ \re{12.7} is written for $qq$-\ia\ in the $SU_f(2)$ flavor sector
of the $SU_f(3)$ theory.
Selection of $SU_f(2)$ corresponds to the case when one of the \q s interacts
with a \vm\ \cd . For the $q\bar q$ - system one should in \re{12.7}
change
\[
\lambda_{\ds q}\rightarrow-\lambda_{\ds\bar q}^T,\ \ \ \ \
\sigma_{\ds q}\rightarrow - \sigma_{\ds \bar q}^T.\]

Let us note that a specific feature of the \ef ive \La\
\re{12.7} is that only the amplitudes with \q s being
in \zfm\ states \ct{tH} are nonzero:
\be
(\vec\sigma_{\ds i}\oplus\vec c_{\ds i})\ \mid\ >=0,
\la{16.19}\ee
where $\vec\sigma_{\ds i}$ is the spin, $\vec c_{\ds i}$ is the color (subgroup
$SU_c(2)$) of an $i$-th \q . This yields nontrivial spin-spin
forces between \zm\ \q s and is especially important for the phenomenology
of multiquark states \ct{DKZ4Q,Jap4Q,DKH}.

The interaction between quarks with long-wave vacuum fields dominates at
distances of an order of the confinement radius ($R_c\sim 1fm$). At
intermediate distances ($\rho _c\sim 0.3\ fm$) the vacuum structure is
	characterized by high-frequency fluctuations approximated by instantons
($\omega _{inst}\sim 1/\rho _c$).

The hadron energy is a sum of the quark kinetic energy and
the energies of interactions
due to one - gluon and one - instanton exchange and interaction with
condensate:
\ba
E=E_{kin}+\Delta E_{OGE}+\Delta E_{inst}+E_{vac},
\la{E}\ea
where
\ba
&&E_{kin}=\frac{1}{R}\sum_{i=1}^{N}[\kappa(m_iR)^2+m_i^2R^2]^{1/2},\la{Ekin}\\
&&E_{vac}=-\sum_{i}^{u,d,s}N_i<0|\bar q_iq_i|0>A_i(m_iR)R^2+...\la{Evac}\\
&&\Delta E_{inst}=-\frac{\rho_c^2}{R^3}\lambda_0
<H|\sum_{i>j}^{u,d,s}
[1+\frac{3}{32}\lambda^a_i\lambda^a_j(1+\vec \sigma_i\vec \sigma_j)]
|H>,\la{Einst}\\
&&\Delta E_{oge}=-\frac{\alpha_s}{R}\lambda_g
<H|\sum_{i>j}^{N}
(\vec \sigma\lambda^a)_i(\vec \sigma\lambda^a)_j|H>,\la{Eoge}
\ea
Here $N_i$ is the number of quarks of definite flavor; $\kappa(m_iR)$ is
the eigenvalue of a lowest energy quark in a spherical cavity bag of
radius $R$; $\vec \sigma_i, \lambda_i^a$ are the spin and color matrices of
the $i$th quark; $\lambda_0\approx 10.085$, $\lambda_g\approx 0.117$,
$A(m_iR)$ are the values of integrals
over cavity wave functions. These \il s and the values of the spin - color
matrix elements over hadron states $|H>$ are calculated in \ct{DK,DKZPi}.

The hadron mass is given by \re{E} with $R_{bag}=R_{equi}$. The equilibrium
radius $R_{equi}$ for each state is determined by minimizing the energy
in zero order in \ia s: $E_0=E_{kin}+E_{vac}$.  It is just this part of the
energy
operator that takes into account the (important) strangeness content of a \hn .
It obviously splits the states in strangeness.

For calculating the spectrum of $4q$ states we use several approximations.
For $\lambda_0,\ \lambda_g$ we neglect the dependence of these integrals on
strange  \q\ mass and compute them at the zero mass of \q s.
Then $\Delta E_{oge}$ and $\Delta E_{inst}$ are diagonal in the basis
of states of total $SU_{csf}(12)$ group.
The corrections related with this approximation are small of second
order and can be taken into account as perturbation.
Dots in \re{Evac}
mean the terms of higher powers of $R$ with the coefficients containing
the condensates of higher dimensions \ct{DK,DKZPi}. The latter are suppressed
by the powers of small ratio
$(\frac{\omega_{vac}}{\omega_q})^2\propto(\frac{1}{\kappa_q})^2<\frac{1}{4}$
and by a small value of corresponding space coordinate integrals.
Moreover, in this work we also neglect the corrections for the spurious center
of mass
motion. There is nothing to prevent one from extending the approach
proposed to terms of order $m_s/E_0$ in $\Delta E_{int}$ and
$<P^2_{cm}>/E_0^2$
as well as to \con s of \vm\ expectation values of the \op\ of higher
dimensions.
It can be believed that a \con\ of this sort will be helpful in a more detailed
study of four - \q\ states.

In the next part we employ the QCD vacuum induced bag model in the spherical
cavity approximation to low - lying $q^2\bar q^2$ mesons. It has been
successful in describing the ground states of light - quark hadrons
\ct{DK,DKZPi}.

The parameters of the model, the strong coupling constant $\alpha _s$,
the effective size of instanton $\rho _c$, the value of quark condensate
$<q\bar q>$, characterizing various
contributions, have been chosen according to the QCD sum rules,
\be
\alpha _s=0.7, <\bar qq>=(-250MeV)^3
\la{par1}\ee  and the
instanton liquid model
\be
\rho _c=2Gev^{-1}.
\la{par2}\ee
They fit the ground state spectrum quite well.
If the interaction with instantons and condensates
is \ten , we can satisfactorily describe both the mass spectra of hadron
ground states and obtain correct values of
$\pi - \rho $, $N - \Delta$, $\Sigma - \Xi$ and $\eta - \eta '$ splittings
\ct{DKZPi}.
\footnote{The same conclusions about the role of instantons in the spin-spin
forces have been done in \ct{Blaske}.}

An advantage of the bag model is the possibility to apply it to the
calculation of $q^2\bar q^2$ meson spectrum without any new parameters.
Previously \ct{DKH,DKZPi} we have applied this model to the problem of
stability
of the six - quark singlet $H$ dihyperon. Now we extend our study on
$q^2\bar q^2$ mesons \ct{DKZ4Q} to the identification of exotic states
in $N\bar N$ reaction.
\chapter{Mass spectrum of $q^2\bar q^2$ mesons in the quark model with QCD
vacuum induced interaction}

\ \ \ \ \ \ \ Pair forces in diquark ($q^2$)
and quarkonium ($q\bar q$) systems occur due to the interaction via
gluon and instanton exchanges. The gluon exchange contribution
due to a small hyperfine coupling constant is, as a rule, not important.
However, the spectrum of $q^2\bar q^2$ mesons depends crucially on \np\
small size \s\ \ia\ between \q s. In the framework of the QCD vacuum
induced model, just this \ia\ is responsible for $\Delta - N$,
$\pi - \rho$ \spl s and permits us to solve $U_A(1)$ problem related
with large $\eta'$ mass.

To calculate the multiquark hadron spectrum in the quark model, we
should construct the physical state basis where the energy is diagonal.
The \op s of kinetic energy, $E_{kin}$, and
the energy of interaction with a quark condensate, $E_{vac}$,
depend only on the quark masses and thus they are diagonal in the basis of
states with a definite number of s-quarks, $N_s$, ("magically mixed states"
or "magic basis", \ct{Jaf}).
However, the \s\ \ia , $\Delta E_{inst}$, lifts the color - spin - flavor
degeneracy and is diagonal in the basis states of the total
$SU_{csf}(12)$ - group with definite color-spin-flavor quantum numbers.

A situation with the construction of the physical state basis for four-quark
systems is analogous in many respects to the one for a
system of $\eta ,\eta '$ mesons.  First, the physical states of these mesons
due to the \s\ annihilation process have no definite number
of s-quarks, second, they do not enter into any irreducible representation
of the flavor group $SU_f(3)$,  but they are a mixture of the singlet -
$\eta _1$ and octet - $\eta _8$ states determined from the
diagonalisation of energy.  As a result, $\eta-\eta'$ are splitted in mass.
In the MIT model there is no reason for the occurrence of a basis different
from
the
magic one. So it is possible there to classify particles with respect to the
$SU_f(3) $ multiplets: $9_f, 36_f, 18_f$. As we shall see below, in the
general case the \s\ interaction that strongly violates the
Okubo - Zweig - Iizuka (OZI)  rule mixes irreducible representations of the
$SU_f(3)$ group and the separation of particles into flavor multiplets
becomes meaningless. It is important that this \s\ - induced process goes not
only for a color octet $q\bar q$ pair (as in the one-gluon annihilation case
\ct{HogHo,Sema}) which is $1/N_c$ suppressed but also for a color singlet
$q\bar q$ pair. As we shall see, the \s\ \ia\ lifts isospin
degeneracy inherent in the MIT model.

As to the spin degrees of freedom, the ground state of the $q^2\bar q^2$ system
can be in $J^p=0^+, 1^+, 2^+$. Let us first consider scalar four - quark
states.

\section{Scalar $q^2\bar q^2$ mesons}

\ \ \ \  We shall construct the basis of $q^2\bar q^2$ states.
A pair of quarks may be in the following color, spin, and
flavor representations:

\be
3_c\otimes3_c=6_c \oplus \bar 3_c , \ \  2_s\otimes2_s=1_s \oplus 3_s ,
\ \  3_f\otimes3_f=6_f \oplus \bar 3_f
\ee
Then from combinations of a diquark and an anti-diquark which are color
singlets
and obey the Pauli principle we shall construct the basis for
$q^2\bar q^2$ states \ct{D},\ct{Do} :

\ba
&&|q^2, \bar q^2>=|(q_c^2, q_s^2, q_f^2), (\bar q_c^2, \bar q_s^2, \bar
q_f^2)>,
     \nn \\
&&|1>_{cs}=|(6_c, 3_s, \bar 3_f), (\bar 6_c, 3_s, 3_f)>, \ \ \ \
|2>_{cs}=|(\bar 3_c, 1_s, \bar 3_f), (3_c, 1_s, 3_f)>, \nn\\
&&|3>_{cs}=|(6_c, 1_s, 6_f), (\bar 6_c, 3_s, \bar 6_f)>, \ \ \ \
|4>_{cs}=|(\bar 3_c, 3_s, 6_f), (3_c, 1_s, \bar 6_f)> \label{y1}
\ea
Taking irreducible flavor representations:
\ba
&&3_f\otimes\bar3_f=9_f=1_f(9)\oplus8_f(9),  \nn \\
&&6_f\otimes\bar6_f=36_f=1_f(36)\oplus8_f(36)\oplus27_f(36)  \label{y2}
\ea
we obtain for scalar $q^2\bar q^2$ mesons ten possible
basis states with a definite flavor:
\ba
|1>_{cs}1_f(9),\ \  |2>_{cs}1_f(9),&&  |3>_{cs}1_f(36),\ \  |4>_{cs}1_f(36),
\nn \\
|1>_{cs}8_f(9),\ \  |2>_{cs}8_f(9),&&  |3>_{cs}8_f(36),\ \
|4>_{cs}8_f(36),  \label{y3} \\
&&|3>_{cs}27_f,\ \ \ \ \ \ \ |4>_{cs}27_f                \nn
\ea

The physical states are
certain superpositions of these states. Thus, in general, in describing
experimental data it is necessary to choose nine mixing angles of
these states which are free parameters. So, of main interest is
the calculation of these parameters within certain dynamic
mechanism leading to a physical basis.

Earlier we have shown that the \s\ - induced \ia\ dominates in the
unitary - multiplet mixing \ct{DK}.
Thus, we shall suggest this \ia\ as the main dynamic mechanism
for the construction of the physical basis.
The kinetic energy and vacuum energy are easier calculated
in the magic basis. It diagonalizes the strange - \q\ number \op\ and
is connected with the $SU_f(3)$ basis by
the known transformations (\ct{Jaf}, Table 10; \ct{Wong}).  The relation
between
the bases allows us to rewrite either
the instanton contribution in the magic basis or the kinetic energy in
the  $SU_f(3)$ basis. Physical states are found by diagonalisation of
the total energy of a hadron.

In Tables. 1,  2,  3, there are the (rounded values of) masses
of $q^2\bar q^2$ mesons and
their decompositions in the magic basis (MIT bag states) obtained in our model.
Here and in the following we use the notation accepted in \ct{Jaf}.  The
exotic states, $E$, mesons not classifiable as flavor octets or singlets,
carry a subscript denoting the (\ps ) flavor channel to which they couple
and are also labelled by their spin parity and the $SU_f(3)$ multiplet in which
they
lie. The cryptoexotics, $C$, (four - \q\ flavor singlets or octets) carry a
subscript
denoting the \ps\ with the same \qn s and a superscript denoting the number
of $s\bar s$ pairs. States at the center of the $SU_f(3)$ weight diagram
have no subscript. The asterisk means a higher mass multiplet.

The results presented in Tables show that this model predicts the same
range of the masses of $q^2\bar q^2$ states as in the MIT model. However, the
magic basis states which diagonalize the energy \op\ in the MIT model
are strongly mixed by the \s\ \ia\ which is diagonal that respect to the total
$SU_{csf}(12)$ group. This essentially changes the decay properties of
four - \q\ mesons.

To analyze the decays of $q^2\bar q^2$ mesons, it is also necessary
to consider the basis made up of the mesonic pair
$(q\bar q)(q\bar q)$ states,  for which the following color -
spin states are possible :

\ba
|M>=|(q\bar q)(q\bar q)>=|((q\bar q)_c, (q\bar q)_s)((q\bar q)_c, (q\bar q)_s)>
\nn \\
|1>_{cs}=|(1_c, 1_s)(1_c, 1_s)>,\ \  |2>=|(1_c, 3_s)(1_c, 3_s)>,  \nn \\
|3>_{cs}=|(8_c, 1_s)(8_c, 1_s)>,\ \  |4>=|(8_c, 3_s)(8_c, 3_s)>.  \label{y4}
\ea

For exotic mesons (  $I>1$ ) in the basis \re{y4}
the values of masses and the recoupling coefficients practically
coincide with the ones presented in \ct{Jaf}. For the remaining
nonexotic $q^2\bar q^2$
mesons the basis \re{y4} has to be supplemented by the flavor recoupling:

\ba
\begin{tabular} {lll}
$(K\bar K)^{I=1}, $&$(\pi \eta _s)^{I=1}, $&$(\pi \eta _0)^{I=1}, $  \\
$(K \pi )^{I=1/2}, $&$ (K \eta _0)^{I=1/2}, $&$ (K \eta _s)^{I=1/2}, $  \\
$(\pi \pi )^{I=0}, $&$ (\eta _0\eta _0), $&$ (K\bar K)^{I=0},  $ \\
$(\eta _0\eta _s),  $&$  (\eta _s\eta _s) $ & \\
$\eta _s=s\bar s, $&$ \eta _0=\frac{1}{\sqrt{2}}(u\bar u+d\bar d). $ & \\
\end{tabular}
\label{y5}
\ea

The states \re{y4}, \re{y5} form together the basis of states with
definite color-spin-flavor content. In Tables 4,  5,  6
the recoupling coefficients of physical
$q^2\bar q^2$ states in the mesonic pair
$|(q\bar q)(q\bar q)>$ states are presented.

  The $q^2\bar q^2$ mesons most probably decay into a pair of $q\bar q$
mesons according to the OZI - superallowed diagram (Fig. 1).
This means that the process is not accompanied by creation or annihilation
of \q - anti\q\ pairs. The decay width is determined by
(see f.i. \ct{Ach a}):
\ba
&&\Gamma_{C_{i}\ra mm'}(s)=\frac{\left| <C_i|mm'>\right| ^2}
{16\pi M_i}F_{mm'}(s),  \nn  \\
&&<C_i|mm'>=g_0 <C_i|mm'>_c <C_i|mm'>_s <C_i|mm'>_f
\label{y6} \ea
where $ <C_i|mm'>_c,  <C_i|mm'>_s,  <C_i|mm'>_f $ are the color, spin and
flavor recoupling coefficients, respectively, which determine the content of
a four-quark state, $C_i$, in terms of a pair of mesons $MM'$ coupled to the
same
total \qn s. They are collected in Tables 4, 5, 6.
As in \ct{Jaf}, here we do not consider the suppressed decays  (that
require the additional creation of a \q\ pair or the gluon exchange).
As a consequence, an $s-$wave $q^2\bar q^2$ meson can only dissociate to
two colorless $s-$wave $q\bar q$ mesons in a relative $s-$wave. The quantity
$F_{m,m'}(s)$ is a phase space integral which takes into account the finite
widths of reaction products; $g_0^2/4\pi$ is the OZI super-allowed
dimensional strong decay constant of a four-\q\ state into a meson-meson pair.
\footnote{
The overall \cef s $g_0$ and $h_0$
below (Eq.\re{y18}) parametrize the coordinate space overlap (vertex)
\il\ between
a four \q\ bag and two meson bags. Instead of calculating them directly,
which is not a good defined procedure within the static cavity \ap , we fix
these \pr s from the \ex\ and predict the branchings.}

\section{Vector and tensor mesons}

As it has been shown in detail in \ct{D},  the spectrum of $J^p=1^+\  q^2\bar
q^2$
mesons has the following features. The main effect of
the instanton interaction is that
unlike the one - gluon exchange interaction, it mixes multiplets $18_f$ with
$\bar{18}_f$ and $9_f$ with $36_f$. Further, the states with different
isotopic spins which are irreducible in masses in the MIT model are splitted
by the \s\ mechanism ($\sim $ 50-200 Mev).

At the same time for tensor mesons, the instanton \ia\ practically does not
change the results of the MIT model. In Tables  7, 8 the color-spin and flavor
recoupling coefficients for
$q^2\bar q^2$ mesons into a pair of vector mesons VV' are presented.

 So, we obtain (Table 1-8) the basis of physical states of scalar,
vector and tensor four - quark mesons and their mass spectrum
within the quark model with QCD vacuum induced quark interaction.
Next sections are devoted to consideration of the $4q$ states as
participants in $N\bar N$ reaction. It should be noted that
in \ct{Ach g}, \ct{Li} the four-quark states were also used in order to
describe the resonance - type signals in the $\gamma \gamma \ra VV'$ reaction.

\chapter{Four-quark states in $N\bar N$ annihilation}

\ \ \ \ The production of $q^2\bar q^2$ mesons can proceed according to
the OZI - allowed diagram of Fig. 2a. In this process
the annihilating quark - antiquark pair  has
vacuum quantum numbers: $^3P_0, J^{pc}=0^{++}$. Then it follows that
the P-wave of $N\bar N$ process will dominate in this diagram.
Due to this fact this reaction may be used as a good filter for
$q^2\bar q^2$ states.
As in the most part of the previous section, here we do not use
any model space coordinate wave \fu s. Instead,  we parametrize the
bag - bag
dynamics by some vertex constants ($g_0,\ h_0$) and pay attention mainly to the
internal \qn\ structure of processes.

There exists also an alternative way of the $q^2\bar q^2$ meson production.
In Fig. 2b, the OZI - \underline{super}allowed diagram that
dominates over the diagram of Fig. 2a is shown. Moreover, the
processes in Fig. 2b may go in relative S-wave with $s-$wave $q\bar q$ mesons
being in the
$J^{pc}=0^{-+}$ or $1^{--}$ states and  $q^2\bar q^2$ meson being
in $0^{++}, 1^{++}$ or $2^{++}$ states. From Fig. 2b  it is seen
that the diquark from
a nucleon with the anti-diquark from an antinucleon produce
the $q^2\bar q^2$ state which may be easily represented as a superposition of
the
known basis states. The remaining quark from the nucleon and antiquark from
the antinucleon may be also easily combined into the usual meson states. Then,
the dynamics of the interaction is reduced to the vertex of Fig. 2c.
The formula for the cross-section of the process of $N\bar N$ annihilation into
mesons is the following:
\be
d\sigma (N\bar N \ra  mm')\sim \left|
\sum_i \frac{<N\bar N |C_i m"><C_i m"|mm'm">}{m_i^2-s-i \sqrt{s}
\Gamma _i(s)}\right| ^2,    \label{y8}
\la{dsig}\ee
where  $C_i$ is a four - quark state,
$m$ is a meson state, $m_i$ and $\Gamma _i(s)$ are the mass and width of $C_i$,
$<C_i m"|mm'm">\sim <C_i|mm'>$.
To calculate \re{dsig} we have to know the matrix elements $<N\bar N|C_i m">$.
First of all, let us consider
the wave function of a nucleon as a $q^2 q$ system :

\ba
&&|N>=|q^2 q>=|(q^2_cq^2_sq^2_f)(q_cq_sq_f)>,      \nn  \\
&&|N>=\frac{1}{\sqrt{2}}(|(\bar 3_c 3_s 6_f)(3_c 2_s 3_f)>+
|(\bar 3_c 1_s 3_f)(3_c 2_s 3_f)>), \label{y9}        \\
&&|\bar N>=\frac{1}{\sqrt{2}}(|(3_c \bar 3_s \bar 6_f)
(\bar 3_c \bar 2_s \bar 3_f)> + |(3_c 1_s 3_f)(\bar 3_c 2_s \bar 3_f)>). \nn
\ea

{}From the group - theoretical point of view the wave function  of the $N\bar
N$
system may be decomposed in a complete set of  $|q^2 \bar q^2 , q\bar q>$
states. From \re{y9} one can easily obtain the expansion:
\footnote{The color decomposition is trivial, so we do not present it here.}

\ba
&&|N\bar N>=\sum_i <N\bar N|(q^2 \bar q^2) (q\bar q)>_i
|(q^2 \bar q^2)_f , (q\bar q)_f)>_i |(q^2 \bar q^2)_s , (q\bar q)_s)>_i=
\label{y10}  \\
&&=\frac{1}{2} (|36_f, 9_f>|9_s, 4_s>+\sqrt{2}|18_f, 9_f>|3_s, 4_s>+
|9_f, 9_f>|1_s, 4_s>),\nn\ea
where we take into account that
$3_s\otimes3_s\equiv9_s=5_s\oplus3_s\oplus1_s$, and so on (see Table 9).

The total spin of the $|(q^2 \bar q^2)_s(q\bar q)_s>$ system is defined by the
total
spin  of $N\bar N$  that can be $J^p=0^-, 1^-(1_s, 3_s)$. In Table  9
we present the spin recoupling coefficients $<N\bar N|(q^2 \bar q^2)(q\bar
q)>_s$.

The flavor recoupling of the $p\bar p$ system into the
$|(q^2 \bar q^2)_f(q\bar q)_f>$ states has the following form:
\ba
\begin{tabular}{rlll}
$|36_f, 9_f>= $&$ 1/3|E_{\pi \pi}^+ \pi ^-> $&$ +1/3|C_{\pi}^+ \pi ^-> $&    \\
 &$ -\frac{2}{3\sqrt{3}}|E_{\pi \pi}^0 \pi ^0> $&$ -1/3|C_{\pi}^0 \pi ^0>
 $&$ -\frac{1}{3\sqrt{6}}|C \pi ^0>$ \\
 &$ +1/3|E_{\pi \pi}^- \pi ^+> $&$ +1/3|C_{\pi}^- \pi ^+> $&  \\
 & &$ +1/3|C_{\pi }\eta > $&$ +\frac{1}{\sqrt{6}}|C \eta >, $  \\
 \end{tabular}    \la{y11}
\ea
\ba
|9_f, 9_f>=1/\sqrt{2}(-|C^0\eta> +|C^0\pi^0>),  \ \ \ \ \
\label{y12}
\ea
\ba
|18_f, 9_f>=1/\sqrt{6}(|\bar C_{\pi}^0(18_f)\eta>
-|\bar C_{\pi}^0(18_f)\pi ^0>)-  \label{y13} \\
-1/\sqrt{3}(|C_{\pi}^+(18_f)\pi ^->+|C_{\pi}^-(\bar 18_f)\pi ^+>),
\ \ \ \ \nn
     \\
\bar C_{\pi}^0(18_f)=1/\sqrt{2}(C_{\pi}^0(18_f)+C_{\pi}^0(\bar 18_f)),
\ \ \ \ \ \ \  \nn \\
\eta=\frac{u\bar u+d\bar d}{\sqrt{2}} ,
\pi ^0=\frac{d\bar d-u\bar u}{\sqrt{2}}.      \ \ \
\nn
\ea

Here $ E_{\pi \pi}, C_\pi $,  C are four-quark states in the
magic basis ,
and the sign of $ E_{\pi \pi}^{\pm}, C_\pi ^{\pm}, \pi ^{\pm}$ means the charge
of the state.

For the $\bar p n$ system we have the expansions:
\ba
\begin{tabular}{rll}
$|36_f, 9_f>= $&$ \sqrt{2/27}|E_{\pi \pi}^0 \pi ^-> $&$ +1/\sqrt{27}|C \pi ^->-
$ \\
 &$-\sqrt{2}/3|E_{\pi \pi}^- \pi ^0>  $&$ +\sqrt{2}/3|C_\pi ^-\eta>+$  \\
 &$+2/3|E_{\pi \pi}^{--} \pi ^+>, $&   \\
\end{tabular} \la{y14}    \\
|9_f, 9_f>=|C^0\pi^->, \ \ \ \ \ \ \ \ \ \ \ \ \ \ \ \ \ \ \ \ \ \ \ \ \ \ \ \
\label{y15} \ea
\ba
\begin{tabular}{rl}
$|18_f, 9_f>=$&$1/\sqrt{6}(-|C_{\pi}^0(18_f)\pi ^->+
|C_{\pi}^0(\bar{18}_f)\pi ^->)$+  \\
 &+$1/\sqrt{3}(|C_{\pi}^-(18_f)\eta >-|C_{\pi}^-(\bar{18}_f)\eta >)- $  \\
 &$-1/\sqrt{3}(|C_{\pi}^-(18_f)\pi ^0>+|C_{\pi}^-(\bar{18}_f)\pi ^0>). $  \\
\end{tabular}
\label{y16}
\ea

The $\bar p n$ system has exactly defined isospin quantum numbers
I=1, $I_z=-1$ and thus in the expressions  \re{y16} the $C_\pi$ states
have definite G-parity.

So we have obtained color, flavor, and spin recoupling of the $N\bar N$ wave
functions
into the $(q^2\bar q^2)(q\bar q)$ wave functions based
only on the group - theoretical considerations.
This allows us to formulate a simple model of the
$(q^2\bar q^2)(q\bar q)$ meson production in the $N\bar N$ annihilation.
Let us proceed to
the discussion of particular examples from meson spectroscopy.

\chapter{The problem of the $a_0(980)$ meson.}

\ \ \ \  The indication of four-quark nature of the $a_0(980)$ meson as the
$C^s_{\pi}(9_f)$ state is has been given in paper \ct{Jaf}. Later on, in ref.
\ct{Ach a}, the description of the $a_0(980)$ meson as a wide resonance
($\Gamma _{a_0 \ra \pi \eta } \sim 300-500 MeV$) has been suggested. In this
case
the narrow peak in the $\pi \eta $ mas spectrum is connected with the threshold
influence of the $K\bar K$ channel. The influence of other $q^2\bar q^2$
states of the MIT-model basis is supposed to be negligible.

However,  in our model,
as a consequence of the instanton mechanism, to describe
the $a_0(980)$ meson, two states with masses
$m_{theor}=1100MeV $ and $m_{theor}=1350MeV $ are essential  (Table 1).
In Fig. 4a the mass spectrum of the channel $\pi
\eta $  in the reaction $K^-p\ra (C(1100)+C(1350))\Sigma ^+\ra (\pi ^-\eta
)\Sigma ^+$ is presented which is defined by the amplitude of
$<K^-K^0 | C(1100)+C(1350)|\pi ^-\eta >$(Fig. 3):

\ba
\frac{d N}{d m_{\pi \eta }}\sim 1/{16\pi} |\frac{<K^-K^0|C(1100)>
<C(1100)|\pi \eta >}{D_c(s)}+ \nn\\
+\frac{<K^-K^0|C(1350)> <C(1350)|\pi \eta >}{D_c(s)}|^2 ,
\ea
where
\be
D_c(s)=m_c^2 - s^2 -i E \Gamma _c(\sqrt{s}) .
\la{denomi}\ee

It follows from Table 1 that both C(1100) and  C(1350) have a
considerable coupling with the states $C_{\pi }^s(9_f)$ and
$C_{\pi }(36_f)$.
The coupling of $36_f$ with vector mesons as compared with $9_f$
is 20 times stronger.
The use of simple considerations of the vector dominance model makes it
possible
to obtain
the experimental value of the width of the decay $a_0(980)\ra (C(
1100)+C(1350))\ra
\gamma \gamma $ \ct{PDG} without additional parameters. If
we put $g_0=16$, in \re{y6}, then we obtain
\be
 \Gamma _{a_0 \ra \gamma \gamma }Br
(a_0 \ra \pi ^0\eta )\approx 0.19 KeV.
\ee

In the experiment \ct{CB} the $a_0(980)$ meson in the $p\bar p\ra a_0\pi,\
a_0\omega,\ a_0\eta$ reactions has been observed. In this connection it is
important to note that in the $N\bar N$ recoupling
the nonet $9_f$ state with isospin $I=1$ (in the MIT model-$C^s_{\pi }$)
is absent which has been previously interpreted as the
$a_0(980)$ meson in ref. \ct{Jaf}, \ct{Ach a}.
It means that $a_0(980)$ can not be observed in the $N\bar N$ annihilation
within the suggestions of refs. \ct{Jaf}, \ct{Ach a}.

In our model the $a_0(980)$ meson is strongly coupled with  the $C_{\pi
}(36_f)$ state which is in the $N\bar N$
decomposition \re{y11}, \re{y14}. In Figs. 4b, 4c the mass
spectrum of the  $\pi \eta $ system in the reactions
$p\bar p\to a_0\omega$, $p\bar p\to a_0\eta$ is shown. It is defined by:
\ba
\frac{d N}{d m_{\pi \eta \omega }}\sim 1/{16\pi }
|\frac{<N\bar N|C(1100)\omega > <C(1100)\omega |\pi \eta \omega
>}{D_c(s)} +    \nn \\
+\frac{<N\bar N|C(1350)\omega > <C(1350)\omega |\pi \eta \omega
>}{D_c(s)} |^2,   \la{y17}
\ea
\ba
&&<N\bar N|C\omega >\sim
1/2(\frac{1}{2\cdot3})_s1/3<C|36_f>h_0,    \la{y18}        \\
&&<C(1100)|36_f>=0.577 , \ \ \ \  <C(1350)|36_f>=0.556,  \la{y19} \\
&&<C\omega |\pi \eta \omega >\approx <C|\pi \eta >  , \ \ \ \ \ \ \ \ \ \ \ \ \
\ \ \ \  \nn \\
&&<C(1100)|\pi \eta >=0.043g_0, \ \ \ \   <C(1350)|\pi \eta >= 0.237g_0.
\la{y20}
\ea
where the coefficients in \re{y19}  are taken from Table 1 and the coefficients
in \re{y20} are from Table  4. The quantity
$g_0^2/4\pi$ is the OZI super-allowed
dimensional strong coupling constant of transition of the $N\bar N$ system
into four-\q\ and $q\bar q$ mesons.
In \re{y18} the unpolarized initial nucleons are used:
\ba
| N\bar N>=\frac{1}{2}(\sqrt{3}\ \underline{3}_s \oplus \underline{1}_s)
\label{y21}
\ea

Using only the recoupling \re{y11}, \re{y14}, we can predict the ratios of
the $a_0(980)$ meson production cross-sections for different reaction
channels (Table 10).

So, in the framework of our approach, a satisfactory description of the
existing experimental data for the $a_0(980)$ production cross sections
and the width of the decay
$a_0(980) \ra \gamma  \gamma $
and predictions for the cross-sections
in the $N\bar N$ annihilation are obtained.
The properties of this meson are unusual owing to its
being described as mixing of two states, C(1100),
C(1350), and its strong coupling with $C^s_{\pi }(9_f)$ and
$C_{\pi }(36_f)$.

In Refs. \ct{WIsgKK}, a very interesting solution has been suggested for the
$a_0(980)-$ and $f_0(980)-$ meson problem. The authors of this Ref. have found
that
\fw\ the potential model these mesons probably correspond to a very weakly
bound
$K\bar K$ molecule. Due to this structure, the model predicts strong
suppression
of the $a_0(980)$ production in the $N\bar N$ reaction.
It would be crucial to check \ex ally the predictions
of the $a_0(980)$ production in the $N\bar N$ reaction
to discriminate between the
approach suggested in \ct{WIsgKK} and that discussed in the present paper.

\chapter{Tensor $q^2\bar q^2$ mesons in $N\bar N$ scattering}

\ \ \ \  The production of $q^2\bar q^2$ tensor mesons is possible only
when it is accompanied by
the vector meson $(N\bar N\ra TV)$. The most  convincing manifestation of
tensor mesons would be the observation of the exotic
meson $E_{\pi \pi }(36_f)$ in the reaction
\ba
\bar pn \ra E^{--}_{\pi \pi}
\rho ^+ \ra (\rho ^- \rho ^-) \rho ^+\ra \pi ^-\pi ^0\pi ^-\pi ^0\pi
^+ \pi ^0
\la{E--}\ea
with the vertex:
\be
<\bar pn|E^{--}\rho ^+>=
1/2\frac{\sqrt{20}}{3\sqrt{3}}2/3h_0 \label{y201}
\ee
and the coupling constant $h_0$.

Let us list other interesting channels of the tensor meson - production:
\ba
&&\bar pn\ra E_{\pi \pi }^-\rho ^0 \ra (\rho
^-\rho ^0)\rho ^0 \ra (\pi ^-\pi ^0\pi ^+\pi ^-)(\pi ^+\pi ^-), \nn \\
&&\bar pp\ra E_{\pi \pi }^-\rho ^+ \ra (\rho ^-\rho ^0)\rho ^+
\label{y22}, \\
&&\bar pp\ra E_{\pi \pi }^+\rho ^- \ra (\rho ^+\rho ^0)\rho ^-, \nn  \\
&&\bar pn\ra C_{\pi }^-\omega  \ra (\rho^-\omega )\omega, \nn \\
&&\bar pp\ra C_{\pi }^-\rho ^+ \ra
(\rho ^-\omega )\rho^+,\label{y23} \\
&&\bar pp\ra C_{\pi }^+\rho
^- \ra (\rho ^+\omega )\rho ^- , \nn  \\
&&\bar pp\ra C_{\pi }^0\omega
\ra (\rho ^0\omega )\omega, \nn \\
&&\bar pp\ra C_{\pi }^0\rho ^0
\ra (\rho ^0\omega )\rho ^0. \nn
\ea

The ratios of the tensor meson production cross - sections
in the $N\bar N$ reaction for different channels
are presented in Table 11.

{}From \re{y18}, \re{y201} and \re{y21}   one can estimate the relative yield
of the $E^{--}$ mesons:
\ba
\frac{\sigma (\bar pp\ra a_0\pi)}{\sigma
(\bar pn\ra E^{--}\rho ^+)} \leq \frac{3}{80}
\label{y24}\ea

The mass spectrum of the $\rho \rho $ system for a reaction like \re{y22}
has the form:
\ba
&&\frac{dN}{dm_{\rho \rho '}}\sim F_{\rho \rho '}/\pi
\left| \frac{<N\bar N|E\rho ><E|VV'>}{m_E^2-s-i
(\sqrt{s}\Gamma (s)+a/2)}\right| ^2,
\label{y25} \\
&&F_{\rho \rho '}(s)=1/\pi ^2\int \limits_{4m_{\pi}^2}^{(\sqrt{s}-2m_{\pi})^2}
dm^2\frac{m\Gamma _{\rho }(m)}{\left| D_{\rho } (m)\right| ^2}
\int \limits_{4m_{\pi}^2}^{(\sqrt{s}-2m_{\pi})^2}dm'^2\frac{m'\Gamma
_{\rho}(m')}
{\left| D_{\rho }(m')\right| ^2}\rho (s, m, m'),  \nn  \\
&&m\Gamma _{\rho }(m)=m_{\rho }\Gamma _{\rho }m_{\rho }/m\left(\frac{q(m)}
{q(m_{\rho })}\right)^3\frac{1+(Rq(m_{\rho })^2}{1+(Rq(m))^2},   \nn \\
&&q(m)=1/2\sqrt{m^2-4m_{\pi}^2},    D_{\rho }(m)=m_{\rho }^2-m^2-im\Gamma _
{\rho }(m),   \nn  \\
&&s=m_{\rho \rho }^2 ,  R=2GeV^{-1} ,  m_{\rho }=0.321, \Gamma _{\rho
}=0.154GeV, \nn \\
&&\Gamma _{E\ra VV'}(s)=\frac{<E|VV'>^2}{16\pi \sqrt{s}}F_{VV'}(s),
\Gamma _E(s)=\sum_{VV'}\Gamma _{EVV'},   \nn  \\
\label{y26} \\
&&<E|\rho \rho >=g_0\frac{1}{\sqrt{3}} . \ \ \ \ \ \ \ \nn
\ea
and the corresponding formula for the $\rho \omega $ mass spectrum of
the system for a reactions \re{y23}
is as follows:
\ba
&&\frac{dN}{dm_{\rho \omega }}\sim F_{\rho }/\pi
\left| \frac{<N\bar N| C_{\pi}V"><C_{\pi}|VV'>}{m_{C_{\pi}}^2-s-
i(\sqrt{s}\Gamma (s)+a/2)}\right| ^2,  \label{y27} \\
&&F_{\rho }(s)=1/\pi ^2\int \limits_{4m_{\pi}^2}^{(\sqrt{s}-2m_{\omega})^2}
dm^2\frac{m\Gamma _{\rho }(m)}
{\left| D_{\rho }(m)\right| ^2}\rho (s, m_{\omega}, m), \nn \\
&&<C_{\pi}|\rho \omega >=g_0\frac{1}{\sqrt{3}}, \nn \\
&&\Gamma _{C_{\pi}\ra \rho \omega }(s)=\frac{<C_{\pi}|\rho \omega >^2}
{16\pi \sqrt{s}}F_{\rho }(s).  \nn
\ea

 The  parameters $m_E, m_{C_{\pi}}, g_0, a$ can be extracted from the data of
the reaction
$\gamma \gamma \ra q^2\bar q^2 \ra \rho \rho ,\rho \omega $.
In paper \ct{Ach g} it has been
shown that $m_E\approx 1.44 GeV, m_{C_{\pi}}=1.4GeV, g_0^2/(4\pi )
\approx  16.4, a\approx  0.65$.
However, a partial wave analysis of the $\gamma \gamma \ra \rho \rho $
reaction can change these results.

As to scalar charged $q^2\bar q^2$ mesons, their production is 20 times
suppressed as compared to that of tensor mesons. It
is enhanced only if it accompanied by a scalar meson
(for example in the reaction $N\bar N\ra E\pi \ra \rho \rho \pi $
\footnote{Some results for these reactions are presented in ref.
\ct{Li NN}, \ct{Ach NN}}).

The situation for neutral mesons is more difficult.
The $N\bar N$ system is strongly coupled
with scalar mesons from the nonet $C^0(9_f)$ and with tensor mesons from
$36_f$ ($E^0,C$). We obtain for neutral $q^2\bar q^2$ mesons
\footnote{The scalar neutral meson $C(36_f)$ is neglected due to its
weak spin coupling.} :
\ba
\bar pn \ra (E_{\pi \pi}^0+C(36_f)+C(9_f))\rho ^- \ra
\left\{   \begin{tabular}{lr}
$(\rho ^+\rho -)\rho ^- $           \\
$(\rho ^0 \rho ^0)\rho ^- $ \ \ \ ,   \\
$(\omega \omega )\rho ^- $          \\
\end{tabular}  \right.   \la{y28}
\\
\bar pp \ra (E_{\pi \pi}^0+C(36_f)+C(9_f))\rho ^0 \ra
\left\{   \begin{tabular}{lr}
$(\rho ^+\rho -)\rho ^0 $          \\
$(\rho^0 \rho ^0) \rho ^0 $   \ \ \ , \\
$(\omega \omega )\rho ^0 $          \\
\end{tabular}  \right.  \la{y29}
\\
\bar pp \ra (C(36_f)+C(9_f))\omega \ra
\left\{   \begin{tabular}{lr}
$(\rho ^+\rho -)\omega  $           \\
$(\rho ^0 \rho ^0) \omega  $  \ \ \ . \\
$(\omega \omega )\omega  $          \\
\end{tabular}  \right.  \la{y30}
\ea

The mass spectrum of the $\rho \rho$ system in reactions \re{y28}, \re{y29} has
the form:
\ba
&&\frac{dN}{dm_{\rho \rho '}}(s)\sim
F_{\rho \rho '}(s)/\pi|< \frac{<N\bar N|E_{\pi \pi}V"><E|\rho \rho >}{D_E(s)}+
\nn\\
&&+\frac{<N\bar N|C(36_f)V"><C(36_f)|\rho \rho >}{D_{C(36_f)}(s)}+  \nn\\
&&+\frac{<N\bar N|C(9_f)V"><C(9_f)|\rho \rho >}{D_{C(9_f)}(s)}| ^2,
\la{30}\ea
where
\ba
<E|\rho \rho >=g_0\frac{1}{\sqrt{3}}
\left\{   \begin{tabular}{ll}
-$1/\sqrt{3}$,   & for $ \rho ^+\rho ^- $   \\
$ \sqrt{2/3}$,   & for $ \rho ^0\rho ^0  $,  \\
\end{tabular}  \right.  \la{31}
\ea

\ba
<C(36_f)|\rho \rho >=g_0\frac{1}{\sqrt{3}}
\left\{   \begin{tabular}{ll}
 $1/\sqrt{6}$,    & for $\rho ^+\rho ^-$  \\
 $1/\sqrt{12}$,   & for $\rho ^0\rho ^0$  \\
 $\sqrt{\frac{3}{4}}$,   & for $\omega \omega $,  \\
\end{tabular}   \right.   \la{32}
\ea

\ba
<C(1350)|\rho \rho >=g_0
\left\{   \begin{tabular}{ll}
 $0.472$,           & for $\rho ^+\rho ^-$  \\
 $0.334$,           & for $\rho ^0\rho ^0$  \\
 $-0.255$,          & for $\omega \omega $,  \\
\end{tabular}   \right.  \la{33}
\ea

\ba
<p\bar p|TV>=h_0\frac{1}{2}\frac{\sqrt{20}}{3\sqrt{3}}
\left\{   \begin{tabular}{ll}
$-\frac{2}{3\sqrt{3}}$,    & for $E^0(36_f)\rho ^0$  \\
 $-\frac{1}{3\sqrt{6}}$,    & for $C(36_f)\rho ^0$  \\
 $+\frac{1}{3\sqrt{6}}$,    & for $C(36_f)\omega$,  \\
\end{tabular}   \right.   \la{34}
\ea

\ba
<p\bar p|SV>=h_0\frac{1}{2}<C(9_f)|C(1350)>
\left\{   \begin{tabular}{ll}
 $\frac{1}{\sqrt{2}}$,    & for $C(9_f)\rho ^0$  \\
 $-\frac{1}{\sqrt{2}}$,   & for $C(9_f)\omega $,  \\
\end{tabular}   \right.   \la{35}
\ea

\ba
<\bar pn|TV>=h_0\frac{1}{2}\frac{\sqrt{20}}{3\sqrt{3}}
\left\{   \begin{tabular}{ll}
 $\frac{2}{\sqrt{27}}$,    & for $E^0(36_f)\rho ^-$  \\
 $\frac{1}{\sqrt{27}}$,    & for $C(36_f)\rho ^-$,  \\
\end{tabular}   \right. \la{36}
\ea

\ba
<p\bar p|SV>=h_0\frac{1}{2}<C(9_f)|C(1350)>
 \begin{tabular}{ll}
 $1$,    & for $C(9_f)\rho ^-$.  \\
\end{tabular}      \la{37}
\ea
The functions $D(s)$ are defined by \re{denomi}.

In reactions \re{y28}-\re{y30} the \ef s of constructive and destructive
interference are essential. They occur due to the possibility of several $4q$
states to be in an intermediate state. This situation looks like the
\ef s in the $\gamma\gamma\to VV$ reaction where a specific behavior of the
invariant mass spectrum in the $\gamma\gamma$ system has been observed
\ct{Exp}.

In the above expressions all numerical
coefficients are taken from Tables 8,9 and Eqs. \re{y10}-\re{y15}.
In \re{33}, \re{35}, \re{37} we have the C(1350) as an example
of the scalar meson.
These formulae provide a set of predictions on the observation of
the tensor mesons in  the $N\bar N$ annihilation.
Thus, we have shown that the $N\bar N$ annihilation is a
good filter for $0^+,2^+$ $q^2\bar q^2$ mesons.
\\

{\bf CONCLUSION}     \\
\vspace{5mm}

     The spectrum of $q^2\bar q^2$ states has been calculated within the model,
that takes into account the interaction between quarks through the
nonperturbative vacuum of QCD. The nonperturbative vacuum of QCD
 is determined by two field components. The first is related to the long -
 wave vacuum fluctuations of quark and gluon fields (the quark and gluon
 condensates). The second one is constructed with the effective t'Hooft
Lagrangian describing the quark interaction through the instantons of small
 sizes. The consideration of these two aspects of the vacuum interaction
has allowed us to formulate the new bag model which gives a good description
 of the hadron spectrum. The spectrum of multiquark states is of particular
 interest. Among a great number of the $q^2\bar q^2$ states one can find
 several states with the same quantum numbers which,
 in general, should be mixed. The presented approach allows us to calculate
 the mixing angles. These calculations give specific predictions for
 the production cross sections and the widths of the
 $q^2\bar q^2$ meson decays.

     A search for the multiquark states usually requires the
features which could distinguish these states from the hybrid,
 glueball and other ones, to be indicated. The analysis of nucleon-antinucleon
annihilation
 into the mesons permits, from our viewpoint, very easy
 identification of most of the four-quark states. In particular, we have
 obtained a satisfactory description of the scalar $a_0(980)$ meson ($N\bar N
 \rightarrow a_0 \omega$ data). The nature of the $a_0(980)$ meson can be
 clarified by measuring the meson - production cross section in
 different annihilation channels $N\bar N \rightarrow a_0\pi$,
 $a_0\omega$, $a_0 \rho$, $a_0 \eta$.

     From our results we expect also a great number of resonances with the
 four-quark structure in different annihilation channels
 into three vector mesons $N\bar N \rightarrow  (VV')V''$ in
 the two vector meson mass spectrum. The cross sections of four-quark meson
 production should satisfy rigorous ratios. This point allows an easy
 separation of $q^2\bar q^2$ mesons from a great number of the observable
particles. \\

{\bf ACKNOWLEDGEMENT}\\
\vspace{5mm}

We would like to express our deep gratitude to N.N.Achasov, B.A.Arbuzov,
T.Barnes, S.B.Gerasimov, H.Hogaasen, I.T.Obuhovski, U.Strohbush
for useful discussions.

\end{sloppypar}

\newpage
Table  1. Masses of $q^2\bar q^2(J^p=0^+)$ mesons and recoupling
coefficients in the magic basis (I=1).

\begin{center}
 \begin{tabular}{||l|r|r|r|r|r|r||}\hline
m, MeV &$C^s_{\pi }(9)$&$C^s_{\pi }(9^*)$& $C_{\pi }(36^*)$&
        $C_{\pi }(36)$&$C^s_{\pi }(36^*)$& $C^s_{\pi }(36)$   \\   \hline
  1100 &  .814  & -.021  &  .025  &  .577  &  -.004 &   .050    \\
  1350 & -.438  & -.034  & -.010  &  .556  &   .014 &   .705    \\
  1700 &  .254  &  .736  &  .084  & -.379  &   .017 &   .493    \\
  1700 & -.014  & -.139  &  .989  & -.030  &   .028 &   .022    \\
  1800 & -.282  &  .661  &  .113  &  .462  &   .028 &  -.506    \\
  2050 &  .014  & -.027  & -.032  & -.012  &   .999 &  -.005    \\  \hline
    \end{tabular}
\end{center}

Table 2. Masses of $q^2\bar q^2(J^p=0^+)$ and recoupling coefficients
in the magic basis (I=1/2).

\begin{center}
\begin{tabular}{||l|r|r|r|r|r|r||} \hline
m, MeV &$C^s_{K}(9)$&$C^s_{K}(9^*)$& $C_{K}(36^*)$&
        $C_{K}(36)$&$C^s_{K}(36^*)$&$C^s_{K}(36)$   \\  \hline
 970 &  .958  &  .000  &  .015  &  .271  &   .008 &  .093     \\
 1400& -.182  & -.133  &  .009  &  .821  &  -.009 & -.524     \\
 1550&  .000  &  .985  & -.034  &  .060  &  -.015 & -.157     \\
 1990&  .000  &  .028  &  .997  & -.038  &  -.040 & -.010     \\
 2000& -.222  &  .104  &  .058  &  .496  &  -.014 &  .831     \\
 2200& -.012  &  .016  &  .040  &  .012  &   .999 &  .002     \\   \hline
  \end{tabular}
\end{center}

Table  3. Masses of $q^2\bar q^2(J^p=0^+)$ and recoupling
coefficients in the basis of the MIT model (I=0).

\begin{center}

 \begin{tabular}{||l|r|r|r|r|r|r|r|r|r|r||} \hline
 m &{\fs C\ (9)}&{\fs C\ $(9^*)$}&{\fs  C\ $(36^*)$}&{\fs C\ (36)}&
{\fs $C^s$\ (9)}&{\fs $C^s\ (9^*)$}&{\fs $C^s\ (36^*)$}&
{\fs $C^s$\ (36)}&{\fs $C^{ss}\ (36^*)$}&{\fs $C^{ss}$\ (36)}  \\
        \hline
 800 & .924&-.003&-.030&-.350&-.045& .007&-.040& .130&-.002& .034     \\
 1140&-.108&-.107&-.039&-.293& .872&-.019&-.020& .311&-.022&-.176     \\
 1350& .040& .978& .084& .034& .150&-.001& .066&-.059& .011&-.062     \\
 1600& .123& .071&-.550& .584& .026& .101&-.267& .485&-.113& .074     \\
 1700& .144&-.077& .787& .443& .063& .137& .013& .367&-.000&-.006     \\
 1700& .045&-.010&-.176& .095& .100& .964&-.016&-.134& .020& .018     \\
 1950& .130&-.080&-.191& .240& .046&-.052& .886& .029& .197&-.230     \\
 2100&-.207& .096&-.012&-.312&-.144& .115& .309& .487&-.043& .694     \\
 2350&-.036& .021&-.025&-.026&-.029&-.005&-.201& .120& .969& .034     \\
 2600& .181&-.075& .021& .297& .419&-.160& .028&-.493& .073& .650 \\ \hline
  \end{tabular}
\end{center}
\newpage
Table  4. Recoupling coefficients for $q^2\bar q^2(J^p=0^+)$ mesons
into the pair of $(q\bar q)(q\bar q)$ mesons  (I=1).

\begin{center}
 \begin{tabular}{|l|c|r|r|r|r|} \hline
m, MeV & f &$|1>_{cs}$&$|2>_{cs}$&$|3>_{cs}$&$|4>_{cs}$      \\ \hline
 1100&$K^+K^-       $& -.288  &  .026  &  .087  &  -.274     \\
     &$K^0K^0       $&  .288  & -.026  & -.087  &   .274     \\
     &$\pi ^0\eta _s$&  .453  & -.029  & -.090  &   .345     \\
     &$\pi ^0\eta _0$&  .373  &  .121  &  .219  &  -.364     \\
     &              &         &        &        &            \\
 1350&$K^+K^-       $&  .387  &  .070  &  .113  &  -.072     \\
     &$K^0K^0       $& -.387  & -.070  & -.113  &   .072     \\
     &$\pi ^0\eta _s$&  .096  &  .093  &  .234  &  -.522     \\
     &$\pi ^0\eta _0$&  .358  &  .091  &  .233  &  -.345     \\
     &              &         &        &        &            \\
 1700&$K^+K^-       $&  .130  & -.182  & -.113  &  -.387     \\
     &$K^0K^0       $& -.130  &  .182  &  .113  &   .387     \\
     &$\pi ^0\eta _s$&  .267  &  .398  &  .428  &   .109     \\
     &$\pi ^0\eta _0$& -.241  & -.001  & -.208  &   .222     \\
     &              &         &        &        &            \\
 1700&$K^+K^-       $&  .001  &  .057  &  .037  &   .024     \\
     &$K^0K^0       $& -.001  & -.057  & -.037  &  -.024     \\
     &$\pi ^0\eta _s$&  .021  & -.045  & -.066  &  -.059     \\
     &$\pi ^0\eta _0$&  .021  &  .730  & -.651  &  -.149     \\
     &              &         &        &        &            \\
 1800&$K^+K^-       $&  .001  & -.253  & -.342  &   .112     \\
     &$K^0K^0       $& -.001  & -.253  &  .342  &  -.112     \\
     &$\pi ^0\eta _0$& -.461  &  .261  &  .167  &   .281     \\
     &$\pi ^0\eta _s$&  .302  &  .166  &  .115  &  -.307     \\
     &              &         &        &        &            \\
 2050&$K^+K^-       $&  .012  &  .380  & -.314  &  -.082     \\
     &$K^0K^0       $& -.012  & -.380  &  .314  &   .082     \\
     &$\pi ^0\eta _s$&  .037  &  .512  & -.471  &  -.119     \\
     &$\pi ^0\eta _0$& -.009  & -.026  &  .016  &   .013     \\
     &              &         &        &        &            \\  \hline
  \end{tabular}
\end{center}
\newpage

Table  5. Recoupling coefficients for $q^2\bar q^2(J^p=0^+)$ mesons
into the pair of $(q\bar q)(q\bar q)$ mesons  (I=1/2).

\begin{center}
 \begin{tabular}{|l|c|r|r|r|r|} \hline
m, MeV & f &$|1>_{cs}$& $ |2>_{cs}$&$|3>_{cs}$&$|4>_{cs}$   \\   \hline
970  &$K^+\pi ^-    $&  .575  & -.004  & -.074  &   .367     \\
     &$K^0\pi ^0    $& -.407  &  .003  &  .052  &  -.260     \\
     &$\pi ^0\eta _0$& -.204  &  .071  &  .168  &  -.458     \\
     &$\pi ^0\eta _s$&  .050  &  .022  &  .033  &  -.059     \\
     &              &         &        &        &            \\
 1400&$K^+\pi ^-    $&  .137  &  .007  &  .097  &  -.331     \\
     &$K^0\pi ^0    $& -.097  & -.005  & -.069  &   .234     \\
     &$\pi ^0\eta _0$&  .514  &  .170  &  .311  &  -.359     \\
     &$\pi ^0\eta _s$& -.338  & -.099  & -.207  &   .328     \\
     &              &         &        &        &            \\
 1550&$K^+\pi ^-    $& -.108  &  .443  &  .453  &   .271     \\
     &$K^0\pi ^0    $&  .076  & -.313  & -.320  &  -.191     \\
     &$\pi ^0\eta _0$&  .119  & -.330  & -.267  &  -.228     \\
     &$\pi ^0\eta _s$& -.102  & -.039  & -.054  &   .100     \\
     &              &         &        &        &            \\
 1900&$K^+\pi ^-    $&  .003  &  .312  & -.257  &  -.051     \\
     &$K^0\pi ^0    $& -.002  & -.221  &  .182  &   .036     \\
     &$\pi ^0\eta_0 $&  .016  &  .626  & -.580  &  -.131     \\
     &$\pi ^0\eta_s $& -.034  & -.039  &  .006  &   .038     \\
     &              &         &        &        &            \\
 2000&$K^+\pi ^-    $&  .002  &  .107  &  .140  &  -.202     \\
     &$K^0\pi ^0    $& -.001  & -.076  & -.099  &   .143     \\
     &$\pi ^0\eta _0$&  .371  &  .075  &  .092  &  -.226     \\
     &$\pi ^0\eta _s$&  .535  &  .136  &  .347  &  -.515     \\
     &              &         &        &        &            \\
 2200&$K^+\pi ^-    $& -.005  &  .021  &  .000  &  -.007     \\
     &$K^0\pi ^0    $&  .003  & -.015  &  .000  &   .005     \\
     &$\pi ^0\eta _0$&  .014  &  .022  & -.025  &  -.012     \\
     &$\pi ^0\eta _s$&  .042  &  .743  & -.645  &  -.170     \\
     &              &         &        &        &            \\ \hline
  \end{tabular}
\end{center}
\newpage

Table  6. Recoupling coefficients for $q^2\bar q^2(J^p=0^+)$  mesons
into the pair of $(q\bar q)(q\bar q)$ mesons  (I=0).

\begin{center}
 \begin{tabular}{|l|c|r|r|r|r|} \hline
m, MeV & f &$|1>_{cs}$& $ |2>_{cs}$&$|3>_{cs}$&$|4>_{cs}$   \\ \hline
800  &\fs  $\pi ^+\pi ^-$  &\fs  .394  &\fs -.063  &\fs -.162  &\fs .513     \\
     &\fs $\pi ^0\pi ^0$  &\fs   .278  &\fs  -.044  &\fs  -.115  &\fs    .363
  \\
     &\fs $K^+K^-        $&\fs   .024  &\fs   .000  &\fs   .045  &\fs   -.050
  \\
     &\fs $K^0K ^0       $&\fs   .024  &\fs   .000  &\fs   .045  &\fs   -.050
  \\
     &\fs $\eta _0\eta _0$&\fs  -.540  &\fs  -.053  &\fs  -.028  &\fs   -.105
  \\
     &\fs $\eta _0\eta _s$&\fs   .083  &\fs  -.009  &\fs   .047  &\fs   -.034
  \\
     &\fs $\eta _s\eta _s$&\fs   .022  &\fs   .004  &\fs   .016  &\fs   -.021
  \\
 1140&\fs $\pi ^+\pi ^-$  &\fs  -.121  &\fs  -.078  &\fs  -.072  &\fs   -.003
  \\
     &\fs $\pi ^0\pi ^0$  &\fs  -.085  &\fs  -.055  &\fs  -.051  &\fs   -.002
  \\
     &\fs $K^+K^-      $  &\fs   .425  &\fs  -.004  &\fs  -.010  &\fs    .183
  \\
     &\fs $K^0K ^0     $  &\fs   .425  &\fs  -.004  &\fs  -.010  &\fs    .183
  \\
     &\fs $\eta _0\eta _0$&\fs  -.134  &\fs  -.038  &\fs  -.057  &\fs    .220
  \\
     &\fs $\eta _0\eta _s$&\fs  -.320  &\fs   .062  &\fs   .211  &\fs   -.528
  \\
     &\fs $\eta _s\eta _s$&\fs  -.114  &\fs   .047  &\fs  -.058  &\fs    .113
  \\
 1350&\fs $\pi^+  \pi^-$  &\fs  -.091  &\fs   .472  &\fs   .409  &\fs    .285
  \\
     &\fs $\pi ^0\pi ^0$  &\fs  -.064  &\fs   .334  &\fs   .290  &\fs    .202
  \\
     &\fs $K^+K^-        $&\fs   .038  &\fs   .016  &\fs  -.047  &\fs    .061
  \\
     &\fs $K^0K ^0       $&\fs   .038  &\fs   .016  &\fs  -.047  &\fs    .061
  \\
     &\fs $\eta _0\eta _0$&\fs   .094  &\fs  -.255  &\fs  -.336  &\fs   -.242
  \\
     &\fs $\eta _0\eta _s$&\fs  -.104  &\fs   .032  &\fs  -.029  &\fs   -.050
  \\
     &\fs $\eta _s\eta _s$&\fs  -.040  &\fs  -.003  &\fs  -.032  &\fs    .037
  \\
 1600&\fs $\pi ^+\pi ^-$  &\fs   .200  &\fs  -.096  &\fs   .259  &\fs   -.034
  \\
     &\fs $\pi ^0\pi ^0$  &\fs   .142  &\fs  -.068  &\fs   .183  &\fs   -.024
  \\
     &\fs $K^+K^-        $&\fs   .169  &\fs  -.090  &\fs   .151  &\fs   -.141
  \\
     &\fs $K^0K ^0       $&\fs   .169  &\fs  -.090  &\fs   .151  &\fs   -.141
  \\
     &\fs $\eta _0\eta _0$&\fs   .267  &\fs  -.285  &\fs   .502  &\fs   -.289
  \\
     &\fs $\eta _0\eta _s$&\fs   .187  &\fs  -.033  &\fs   .309  &\fs   -.165
  \\
     &\fs $\eta _s\eta _s$&\fs   .043  &\fs  -.071  &\fs   .104  &\fs   -.027
  \\
 1700&\fs $\pi ^+\pi ^-$  &\fs   .215  &\fs   .231  &\fs  -.185  &\fs   -.123
  \\
     &\fs $\pi ^0\pi ^0$  &\fs   .152  &\fs   .164  &\fs  -.131  &\fs   -.087
  \\
     &\fs $K^+K^-        $&\fs   .130  &\fs   .080  &\fs   .108  &\fs   -.067
  \\
     &\fs $K^0K ^0       $&\fs   .130  &\fs   .080  &\fs   .108  &\fs   -.067
  \\
     &\fs $\eta _0\eta _0$&\fs   .215  &\fs   .602  &\fs  -.248  &\fs   -.385
  \\
     &\fs $\eta _0\eta _s$&\fs   .152  &\fs  -.008  &\fs   .047  &\fs   -.232
  \\
     &\fs $\eta _s\eta _s$&\fs  -.004  &\fs  -.001  &\fs  -.002  &\fs    .004
  \\
     &                &        &        &        &            \\ \hline
 \end{tabular}

 \begin{tabular}{|l|c|r|r|r|r|} \hline
m, MeV & f &$|1>_{cs}$& $ |2>_{cs}$&$|3>_{cs}$&$|4>_{cs}$   \\ \hline
 1700&$\pi ^+\pi ^-$  &  .047  & -.053  &  .052  &   .006     \\
     &$\pi ^0\pi ^0$  &  .033  & -.037  &  .037  &   .004     \\
     &$K^+K^-        $& -.092  &  .291  &  .270  &   .272     \\
     &$K^0K ^0       $& -.092  &  .291  &  .270  &   .272     \\
     &$\eta _0\eta _0$&  .029  & -.094  &  .139  &  -.038     \\
     &$\eta _0\eta _s$&  .006  & -.461  & -.444  &  -.262     \\
     &$\eta _s\eta _s$&  .013  &  .010  & -.006  &  -.015     \\
     &                &        &        &        &            \\
 1950&$\pi ^+\pi ^-$  &  .138  & -.081  &  .040  &  -.011     \\
     &$\pi ^0\pi ^0$  &  .098  & -.057  &  .028  &  -.008     \\
     &$K^+K^-        $&  .049  &  .314  & -.301  &  -.079     \\
     &$K^0K ^0       $&  .049  &  .314  & -.301  &  -.079     \\
     &$\eta _0\eta _0$&  .072  & -.058  &  .227  &  -.127     \\
     &$\eta _0\eta _s$&  .008  &  .494  & -.368  &  -.125     \\
     &$\eta _s\eta _s$& -.140  &  .106  & -.221  &   .110     \\
     &                &        &        &        &            \\
 2100&$\pi ^+\pi ^-$  & -.203  &  .023  &  .018  &   .013     \\
     &$\pi ^0\pi ^0$  & -.144  &  .017  &  .013  &   .009     \\
     &$K^+K^-        $&  .099  &  .198  &  .047  &  -.201     \\
     &$K^0K ^0       $&  .099  &  .198  &  .047  &  -.201     \\
     &$\eta _0\eta _0$& -.089  & -.091  & -.151  &   .218     \\
     &$\eta _0\eta _s$&  .321  &  .167  & -.069  &  -.218     \\
     &$\eta _s\eta _s$&  .445  &  .091  &  .310  &  -.425     \\
     &                &        &        &        &            \\
 2350&$\pi ^+\pi ^-$  & -.029  &  .001  &  .016  &  -.002     \\
     &$\pi ^0\pi ^0$  & -.021  &  .001  &  .011  &  -.001     \\
     &$K^+K^-        $&  .024  & -.065  &  .090  &  -.031     \\
     &$K^0K ^0       $&  .024  & -.065  &  .090  &  -.031     \\
     &$\eta _0\eta _0$& -.000  & -.028  & -.005  &   .025     \\
     &$\eta _0\eta _s$&  .054  & -.089  &  .125  &  -.011     \\
     &$\eta _s\eta _s$&  .061  &  .726  & -.613  &  -.185     \\
     &                &        &        &        &            \\
 2600&$\pi ^+\pi ^-$  &  .183  & -.012  & -.011  &  -.016     \\
     &$\pi ^0\pi ^0$  &  .130  & -.008  & -.008  &  -.011     \\
     &$K^+K^-        $&  .012  & -.093  & -.195  &   .254     \\
     &$K^0K ^0       $&  .012  & -.093  & -.195  &   .254     \\
     &$\eta _0\eta _0$&  .093  &  .087  &  .132  &  -.207     \\
     &$\eta _0\eta _s$& -.454  &  .038  & -.034  &   .068     \\
     &$\eta _s\eta _s$&  .422  &  .169  &  .217  &  -.417     \\
\hline
  \end{tabular}
\end{center}
\newpage
Table  7. Recoupling coefficients for $q^2\bar q^2$ mesons into pair of
vector color - singlet VV' and color-octet $\underline{VV'}$ states.

\begin{center}
\begin{tabular}{|r|rr|} \hline
 & VV' & $\underline{VV'}$ \\  \hline
        $9_f$ & $ \sqrt{2/3} $ &-$\sqrt{1/3} $   \\
$36_f$ & $\sqrt{1/3}$ & $\sqrt{2/3}$  \\  \hline
\end{tabular}
\end{center}

Table  8. Recoupling coefficients in flavor for tensor mesons into vector
mesons.

\begin{center}

\begin{tabular}{|r|ccccccccc|} \hline

flavor st.&$\rho ^+\rho ^-$&$\rho
^0\rho ^0$&$K^{*+}K^{*-}$&$K^{*0} \bar K^{*0}$&$\omega \omega $&$\omega
\phi $&$\rho ^0\phi$&$\rho ^0 \omega$&$\phi \phi $  \\     \hline

$C^0(9_f)$ &$\sqrt{1/2}$&1/2& 0 & 0 &-1/2& 0 & 0 & 0 & 0 \\

$C^s(9_f)$ & 0 & 0 &1/2&1/2& 0 &-$\sqrt{1/2}$& 0 & 0 & 0 \\ $C^s_{\pi}
(9_f)$ & 0 & 0 &-1/2&1/2& 0 & 0 &$\sqrt{1/2}$& 0 & 0 \\

$E(36_f)$ &-$\sqrt{1/3}$& $\sqrt{2/3}$&0&0&0&0&0&0&0 \\

$C^0_{\pi }(36_f)$& 0 & 0 & 0 & 0 & 0 & 0 & 0 & 1 & 0 \\

$C^0(36_f)$ &$\sqrt{1/6}$&$\sqrt{1/12}$&0&0&$\sqrt{3/4}$&0&0&0&0 \\

$C^s_{\pi }(36_f)$& 0 & 0 &1/2&-1/2& 0 & 0 &$\sqrt{1/2}$& 0 & 0 \\

$C^s(36_f)$ & 0 & 0 &1/2&1/2& 0 &$\sqrt{1/2}$& 0 & 0 & 0 \\

$C^{ss}(36_f)$    & 0 & 0 & 0 & 0 & 0 & 0 & 0 & 0 & 1  \\  \hline
\end{tabular}
\end{center}
Table  9. The spin recoupling coefficients $<N\bar N|(q^2 \bar q^2)(q\bar
q)>_s$.

\begin{center}
\begin{tabular}{|lc|rrrrr|} \hline
                &  & TV & SV & VS & SS & VV   \\
$J^{pc}_{total}$ & $(q^2 \bar q^2 , q\bar q)$  &$ 2^+1^-$ &$ 0^+1^-$&$1^+0^-$
&$ 0^+0^-$ &$ 1^+1^-$    \\   \hline
 $1^-(3_s)$ & $(9_s, 4_s)$ &$\frac{\sqrt{20}}{3\sqrt{3}}$&$-\frac{1}
{3\sqrt{3}}$&$\frac{\sqrt{2}}{3}$& 0 & 0  \\
 $0^-(1_s)$ & $(9_s, 4_s)$ & 0 & 0 & 0 &$\frac{1}{\sqrt{3}}$ &
       $\sqrt{\frac{2}{3}}$  \\
        $1^-(3_s)$ & $(3_s, 4_s)$ & 0 & 0 &$ \frac{1}
{\sqrt{3}}$& 0 &$ \sqrt{\frac{2}{3}}$  \\
 $0^-(1_s)$ & $(3_s, 4_s)$ & 0 & 0 & 0 & 0 & 1 \\
 $1^-(3_s)$ & $(1_s, 4_s)$ & 0 & 1 & 0 & 0 & 0 \\
 $0^-(1_s)$ & $(1_s, 4_s)$ & 0 & 0 & 0 & 1 & 0 \\        \hline
        \end{tabular}
\end{center}
Table 10. The ratios of the $a_0(980)$ meson production cross - sections
in $N\bar N$ reaction for different channels,  $cos^2\theta \sim 2/3$.
$\theta$ is a mixing angle in $\eta\eta'$ sysytem.

\begin{center}
 \begin{tabular}{|l|c|c|c|c|c|c||} \hline
 & $p\bar p \ra a_0\rho $ &
  $p\bar p \ra a_0\eta $ & $p\bar p \ra a_0\omega $ &
$\bar pn \ra a_0^-\eta $ & $\bar pn \ra a_0^-\omega $    \\
                                \hline
$\frac{\sigma (N\bar N\ra a_0x)}{\sigma (p\bar p\ra a_0\pi )} $ &
 1/3 & $cos^2\theta $ & 1/3 & $2cos^2\theta $ & 2/3      \\
                        \hline
\end{tabular}
\end{center}
\newpage

Table 11. The ratios of the tensor meson production cross - sections
in $N\bar N$ reaction for different channels.

\begin{center}
\begin{tabular}{|l|c|l|c|} \hline
&$ \frac{\sigma (\bar NN\ra TV)}{\sigma (\bar pn\ra E^{--}\rho ^+)} $& &
$ \frac{\sigma (\bar NN\ra TV)}{\sigma (\bar pn\ra E^{--}\rho ^+)} $ \\
\hline
$\bar pn\ra E^{-}\rho ^0 $& 1/2 &$
\bar pn\ra C^{-}_{\pi } \omega  $& 1/2 \\
$\bar pp\ra E^{-}\rho ^+ $& 1/4 &$
\bar pp\ra C^{-}_{\pi } \rho ^+ $& 1/4 \\
$\bar pp\ra E^{+}\rho ^- $& 1/4 &$
\bar pp\ra C^{+}_{\pi } \rho ^- $& 1/4 \\
$\bar pp\ra C^{0}_{\pi } \rho ^0$& 1/4 &$
\bar pp\ra C^{0}_{\pi } \omega  $& 1/4 \\
\hline
\end{tabular}
\end{center}

\vspace{25mm}

LIST OF CAPTIONS.\\

Figure 1. The diagrams of a) $q^2\bar q^2$ meson decay and b)
$q^2\bar q^2$ meson production in $\gamma\gamma$ scattering.\\

Figure 2.
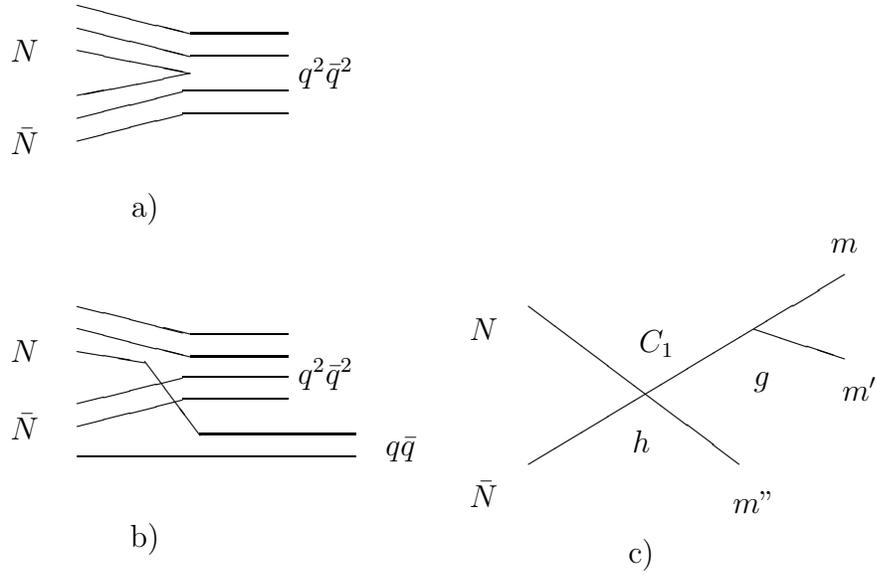
The diagrams of $q^2\bar q^2$ meson production in $N\bar N$ annihilation
a) the OZI - allowed diagram, b) the OZI - \underline{super}allowed diagram,
c) the vertex corresponding to b).\\

Figure 3. The diagram of $a_0(980)$-meson production in $pK^-$ scattering.\\

Figure 4. The mass spectrum of the $\pi\eta $  system in reactions
a) $K^-p\ra (C(1100)+C(1350))\Sigma ^+$ $\ra (\pi ^-\eta)\Sigma ^+$ and
b)$p\bar p\to a_0\omega$, $p\bar p\to a_0\eta$.
\newpage
\unitlength=1.00mm
\special{em:linewidth 0.4pt}
\linethickness{0.4pt}
\begin{picture}(134.00,149.00)
\put(30.00,144.00){\line(1,0){15.00}}
\put(45.00,144.00){\line(2,1){10.00}}
\put(30.00,141.00){\line(1,0){15.00}}
\put(45.00,141.00){\line(2,1){10.00}}
\put(30.00,138.00){\line(1,0){15.00}}
\put(45.00,138.00){\line(2,-1){10.00}}
\put(30.00,135.00){\line(1,0){15.00}}
\put(45.00,135.00){\line(2,-1){10.00}}
\put(23.00,139.00){\makebox(0,0)[cc]{$C$}}
\put(75.00,139.00){\line(1,0){15.00}}
\put(90.00,139.00){\line(2,1){10.00}}
\put(90.00,139.00){\line(2,-1){10.00}}
\put(70.00,139.00){\makebox(0,0)[cc]{$C$}}
\put(67.00,115.00){\makebox(0,0)[cc]{Figure 1. The diagrams of $q^2\bar q^2$
meson decay.}}
\put(30.00,95.00){\line(5,-1){15.00}}
\put(45.00,92.00){\line(-5,-1){15.00}}
\put(30.00,98.00){\line(4,-1){15.00}}
\put(45.00,94.33){\line(1,0){13.00}}
\put(30.00,86.00){\line(4,1){14.00}}
\put(44.00,89.67){\line(1,0){14.00}}
\put(30.00,101.00){\line(4,-1){15.00}}
\put(45.00,97.33){\line(1,0){13.00}}
\put(30.00,83.00){\line(4,1){14.00}}
\put(44.00,86.67){\line(1,0){14.00}}
\put(23.00,95.00){\makebox(0,0)[cc]{$N$}}
\put(23.00,83.00){\makebox(0,0)[cc]{$\bar N$}}
\put(63.00,92.00){\makebox(0,0)[cc]{$q^2\bar q^2$}}
\put(30.00,58.00){\line(4,-1){15.00}}
\put(45.00,54.33){\line(1,0){13.00}}
\put(30.00,48.00){\line(4,1){14.00}}
\put(44.00,51.67){\line(1,0){14.00}}
\put(30.00,61.00){\line(4,-1){15.00}}
\put(45.00,57.33){\line(1,0){13.00}}
\put(30.00,45.00){\line(4,1){14.00}}
\put(44.00,48.67){\line(1,0){14.00}}
\put(23.00,55.00){\makebox(0,0)[cc]{$N$}}
\put(23.00,45.00){\makebox(0,0)[cc]{$\bar N$}}
\put(63.00,52.00){\makebox(0,0)[cc]{$q^2\bar q^2$}}
\put(30.00,41.00){\line(1,0){37.00}}
\put(30.00,55.00){\line(6,-1){9.00}}
\put(39.00,53.67){\line(3,-4){7.33}}
\put(46.33,44.00){\line(1,0){20.67}}
\put(73.00,42.00){\makebox(0,0)[cc]{$q\bar q$}}
\put(39.00,74.00){\makebox(0,0)[cc]{a)}}
\put(39.00,30.00){\makebox(0,0)[cc]{b)}}
\put(90.00,61.00){\line(4,-3){28.00}}
\put(90.00,40.00){\line(5,3){42.00}}
\put(120.00,58.00){\line(3,-1){12.00}}
\put(84.00,58.00){\makebox(0,0)[cc]{$N$}}
\put(84.00,36.00){\makebox(0,0)[cc]{$\bar N$}}
\put(105.00,43.00){\makebox(0,0)[cc]{$h$}}
\put(120.00,35.00){\makebox(0,0)[cc]{$m"$}}
\put(134.00,50.00){\makebox(0,0)[cc]{$m'$}}
\put(121.00,51.00){\makebox(0,0)[cc]{$g$}}
\put(107.00,56.00){\makebox(0,0)[cc]{$C_1$}}
\put(132.00,69.00){\makebox(0,0)[cc]{$m$}}
\put(105.00,28.00){\makebox(0,0)[cc]{c)}}
\put(76.00,8.00){\makebox(0,0)[cc]{Figure 2. The diagrams of $q^2\bar q^2$
meson production in $N\bar N$ annihilation}}
\put(76.00,1.00){\makebox(0,0)[cc]{a) the OZI - allowed diagram, b) the OZI -
superallowed diagram, c) the vertex}}
\put(30.00,-7.00){\makebox(0,0)[cc]{corresponding to b)}}
\end{picture}
\newpage
\unitlength=1.00mm
\linethickness{0.3mm}
\begin{picture}(89.00,142.00)
\put(40.00,135.00){\line(1,0){20.00}}
\put(60.00,135.00){\line(3,1){15.00}}
\put(61.00,135.00){\line(0,-1){15.00}}
\put(41.00,120.00){\line(1,0){34.00}}
\put(75.00,120.00){\line(5,3){9.00}}
\put(75.00,120.00){\line(5,-3){9.00}}
\put(36.00,135.00){\makebox(0,0)[cc]{$P$}}
\put(36.00,119.00){\makebox(0,0)[cc]{$K^-$}}
\put(58.00,128.00){\makebox(0,0)[cc]{$K$}}
\put(79.00,142.00){\makebox(0,0)[cc]{$\Sigma^+$}}
\put(61.00,114.00){\makebox(0,0)[cc]{$g$}}
\put(68.00,116.00){\makebox(0,0)[cc]{$C$}}
\put(75.00,114.00){\makebox(0,0)[cc]{$g$}}
\put(89.00,125.00){\makebox(0,0)[cc]{$\pi^-$}}
\put(89.00,115.00){\makebox(0,0)[cc]{$\eta$}}
\put(5.00,94.00){\makebox(0,0)[lc]{Figure 3. The diagram of $a_0(980)-$ meson
production in $pK^-$ scattering.}}
\end{picture}
\end{document}